\documentstyle{mn}
\begin{document}

\input epsf  

\title[triplets of galaxies]{Dynamical evolution of triplets of galaxies}

\author[Aceves]{H\'ector Aceves \\
Instituto de Astronom\'{\i}a, UNAM. Apdo. Postal 877, Ensenada B.C. 22860, M\'exico. \quad aceves@astrosen.unam.mx 
}
\date{Accepted -------. Received -------; in original
	form -------}

\pagerange{\pageref{firstpage} - - \pageref{lastpage} }
\pubyear{2001}

\maketitle
\label{firstpage}

\begin{abstract}
	 By means of $N$-body simulations we study the global dynamics of triplets of galaxies, considering initial conditions starting from `maximum expansion' and in virial equilibrium.  
	Unlike previous studies we treat galaxies self-consistently, but we
restrict ourselves to models with spherical symmetry and do not consider the influence of a primordial common halo of dark matter.
 
	Our results indicate that a low number of triple mergers are expected at the present epoch ($\approx 10$\%) for collapsing triplets. Initially virialised conditions yield $\approx 5$ percent of triple mergers in $\sim 10$ Gyr of evolution; hence, the 3-galaxy problem has stable states. No overmerging problem  for these small groups of galaxies is found. 
	Their geometrical properties, as reflected by the Agekyan-Anosova map, do not show an excess of extreme hierarchical triplets.
	Unlike the 3-body problem no `sling-shot' events are found during triple interactions, both for collapsing and virial initial conditions.
 	The median velocity dispersion of observed compact triplets ($\sigma \sim 100$ km s$^{-1}$) is not well reproduced in our models at the present epoch: $\sigma \sim 50$ km s$^{-1}$ for collapsing and  $\sigma \sim 80$ km s$^{-1}$ for virial.  
	However,  about 10 per cent of simulated triplets reaching the present epoch from maximum expansion have dynamical properties very similar to the median of Karachentev's compact triplets.  Our median values agree, however, very well with new data on triplets.
	We find that the median of the virial mass estimates do not overestimate, in general, the mass of triplets, but underestimate it by $\approx 35$ percent. The median mass estimator appears as a somewhat better mass estimator. 

	Analysis of the dynamical parameters, as well as  information obtained from a pseudo phase-plane constructed using their velocity dispersion and harmonic radius,  lead us to conclude that:  Karachentsev's compact triplets probably represent the most advanced stage of gravitational clustering of initially diffuse triplets. To test this thesis  we suggest that triplets be studied within a cosmological scenario.

\end{abstract}


\section{Introduction}

	Galaxies in the universe tend to aggregate in  different levels of structure, from binaries to large super-clusters. Systems of three galaxies, or triplets, constitute the smallest galaxy groups and are the boundary of $N$-body systems that cannot be modeled by analytical methods. 
	On other hand, existing compact triplets of galaxies (Karachentsev 2000) are not very distinct from compact groups  of galaxies (Hickson 1997) and  their dynamical study may help in our understanding of the latter.

	Karachentseva, Karachentsev \& Shcherbanovskii~\shortcite{karaa79} were the first to compile a systematic catalog of triplets  and to study some of their observational properties. At the present time,  observational surveys in search of triplets and studies to determine their kinematical properties are being conducted by several groups (e.g., Karachentseva \& Karachantsev 2000; Infante et al. 2000). 
	On other hand, Heidt et al. \shortcite{heidt99} have found that a strongly interacting triplet may be responsible for the production of a  BL Lac object, and  Iglesias-P\'aramo \& V\'{\i}lchez  (1997) show an example of a probable triple interaction of galaxies in Hickson's compact group HCG 95. All this observational research points toward the necessity of developing a better understanding of the dynamics of triplets.

	Triplets maintain also  a close relation with the 3-body problem of celestial mechanics and, hence, their study is also of theoretical interest (Valtonen \& Mikkola 1991; Hein\"am\"aki et al. 1998; Muzzio, Wachlin \& Carpintero 2000). A popular idea in the community (Valtonen \& Flynn 2000) is that there are no stable states in a triplet, but as we will show later this is incorrect.

	In general, the problem of triplets, or 3-galaxy problem, has not received much attention in contrast to the dynamics of compact groups of galaxies. Several dynamical studies of triplets have been carried out by different authors in order to explain some of their observational properties, but  most of the up-to-date studies have used either a point-particle approach to model galaxy interactions (e.g., Chernin \& Mikkola 1991) or included Chandrasekhar's formula to model dynamical friction (Zheng et al. 1993).

 	In the 3-body simulations made by Chernin \& Mikkola (1991) they found too many hierarchical structures (i.e. a close binary and a far away third body) that does not correspond to what is observed.
 	Chernin et al. \shortcite{chernin-aa} have suggested that dark matter may be the necessary ingredient to help 3-body simulations better reproduce  these observations, based up on numerical experiments that show that the probability of binary formation is reduced if dark matter is present in the system \cite{chernin89}.  
	On other hand, Zheng et al. (1993) have shown that the excess of hierarchical structures may be reduced somewhat by the inclusion of Chandrasekhar's dynamical friction formula in simulations of point-like  galaxies; it is to note that Zheng et al. start their simulations with five `galaxies'.

	The use of non self-consistent galaxies in  previous studies lead   to consider current results on the dynamics of triplets as a first approximation to the 3-galaxy problem. 
	 Effects like dynamical friction are not easy to model by an explicit-physics approach, which in any event requires fine tuning using self-consistent galaxies (Vel\'azquez \& White 1999).  Therefore some doubts are cast on the results obtained when the self-gravity of galaxies is not taken into consideration.	The natural following step in the modeling of triplets is to treat galaxies as self-gravitating systems, which will allow them to absorb orbital energy and angular momentum during an encounter.

	An important quantity to  determine in triplets, as well as in other systems of galaxies, is their gravitational mass. 
	Although large amounts of dark matter are rather well established  in groups and clusters of galaxies \cite{neta99}, the situation in low-multiplicity systems is still a matter of some debate.
	For example, several authors have suggested that binary galaxies, even with the known uncertainties in their orbital parameters~\cite{bt}, do not have much more dark matter than that probably associated with their individual haloes (e.g., Karachentsev 1990; Honma 1999).

	On other hand, numerical simulations using point-particles indicate that the mass of a triplet obtained by the virial mass estimator is unreliable, with overestimates up to one or two orders of magnitude (e.g., Kiseleva \& Chernin 1988; Chernin \& Mikkola 1991). 
	Dolgachev \& Chernin \shortcite{dolgachev}, carrying out simulations of wide-triplets (W-triplets, see $\S 2$) with free-fall initial conditions,  conclude that these systems must hold a considerable amount of dark matter and that their masses are significantly underestimated if their non-stationary state is not taken into consideration.
	The assessment of typical methods used to determine mass, when using self-consistent galaxies and involving a non-stationary situation, may shed light towards understanding some of these discrepancies. It will be evident later that the self-gravity of galaxies introduces another degree of uncertainty in the determination of mass due to the transfer of orbital to internal energy during interactions.

	In this contribution we perform a set of numerical experiments on the dynamics of triplets using a self-gravitating  galaxy model. In particular, we will focus on global quantities that are directly comparable with observations in order to test our model.
	We divide this work as follows. In $\S 2$ we present an outline of  some of the observational properties of compact and wide triplets, and our adopted set of definitions to estimate the dynamical quantities. In $\S 3$ the galaxy model and the initial conditions to be used in the simulations are presented. In $\S 4$ our results are presented and discussed.  Here we consider the general aspects of the evolution of triplets, merging histories, geometrical properties, and the median values of their observationally related dynamical quantities. Finally, in $\S 5$ we summarize the main results of this work.

\section{Observational setting}

Karachentseva et al.~\shortcite{karaa79} have presented the first systematic study of the properties of triplets having a compact configuration in the sky (see also Karachentsev 2000; Karachentseva \& Karachentsev 2000).
 	According to the statistical criteria of Anosova \shortcite{anosova87} to distinguish physical from optical triplets, which has become a standard in selecting physical triplets and is considered to be a strong selection criteria (Zheng et al. 1993), about 45 triplets out of 83 are probably physical; we will refer to the former as K-triplets or compact triplets.

	Trofimov \& Chernin \shortcite{trofimov} have analyzed the triplets in the catalogs of Maia, da Costa \& Latham (1989) and Huchra \& Geller (1982).     For these triplets, they derive a median mean harmonic radius that is  about an order of magnitude larger than that of K-triplets, thus calling them  wide triplets (W-triplets). 
	In Table \ref{tab:triplets} the median values of different dynamical quantities, both for K-triplets and W-triplets, are provided for future reference.  The quantities written are: the mean harmonic radius, $R_{\rm H}$; the one-dimensional velocity dispersion, $\sigma$; the dimensionless crossing time, $H_0 \tau_{\rm c}$, and the virial mass estimate $M_{\rm v}$. A Hubble constant of $H_0=75$ km s$^{-1}$ Mpc$^{-1}$ is assumed throughout this work.

\begin{table}
\caption[]{Median dynamical properties of triplets}
\label{tab:triplets}
\begin{flushleft}
\begin{tabular}{lrccc}
\hline\noalign{\smallskip}
     &  $R_{\rm H}$   &  $\sigma$  & $H_0 \tau_{\rm c}$ & $M_{\rm v}$  \\
     &  [kpc]   &  [km/s] &  & [M$_\odot$]  \\
\noalign{\smallskip}
\hline
\noalign{\smallskip}
K-Triplets   & 65.7  & 120.0 & 0.041  & $1.71 \times 10^{12}$ \\
W-Triplets   & 653.5 & 105.3 & 0.531  & $9.46 \times 10^{12}$ \\
\noalign{\smallskip}
\hline
\end{tabular}    
\end{flushleft}
\end{table}

	The values for K-triplets in Table~\ref{tab:triplets} were obtained using the information for $\sigma$ and $R_{\rm H}$ provided by Karachentsev et al. (1989), and for   W-triplets' those  found in Trofimov \& Chernin (1995) and Dolgachev \& Chernin (1997). 
 	The differences of values in Table~\ref{tab:triplets} with those found in the referred works arise from a difference in the definitions used for the dynamical parameters.  
	The expressions used here, as well as that for the median mass estimator to be used later, are as follows (e.g., Nolthenius \& White 1987; Heisler, Tremaine \& Bahcall 1987):
\begin{eqnarray} \label{eq:rh}
\frac{1}{R_{\rm H}}  & = & \frac{2}{N (N-1)} \frac{2}{\pi} \sum_{i<j} \frac{1}{R_{ij}} \,, \\
\label{eq:sigma}
\sigma^2 & = & \frac{1}{N-1} \sum_i \; ( v_i - \langle v \rangle )^2 \,, \\
\label{eq:tcross}
\tau_{\rm c} &=& \frac{2\, R_{\rm H}}{\sqrt{3}\,\sigma} \,,
\end{eqnarray}
where $\langle v \rangle = N^{-1} \sum v_i \,$, $N=3$,  and 
\begin{equation} \label{eq:mvirial}
M_{\rm v} = \frac{6}{G} \, R_{\rm H}\, \sigma^2 \,, \;\;
M_{\rm med} = \frac{6.5}{G} \, {\rm MED}_{ij}\, [ (v_i - v_j)^2 R_{ij} ]\,,
\end{equation}
with $v_i$ being the line-of-sight velocity of a galaxy and $R_{ij}$ the projected pair-wise separation of two galaxies. In these formulae the bulk velocity of galaxies and their position vectors are assumed to be randomly oriented.

	The expression for the virial mass yields the standard formula of the virial mass estimator, which leads to accurate results for the total mass of an  equilibrium system made  of point-like particles (e.g., Aceves \& Perea 1999).  The median mass estimator is used here due to its insensitivity to extreme values,  and also to compare its performance against the virial mass estimator.
	 In the expression for $M_{\rm med}$ a Michie's equilibrium model was used to obtain the constant factor;  although this factor is perhaps not true for triplets, we none the less use it as is normal practice and also to assess its accuracy in a triplet of self-consistent galaxies.

\begin{figure} 
\epsfxsize=8.5cm
\epsffile{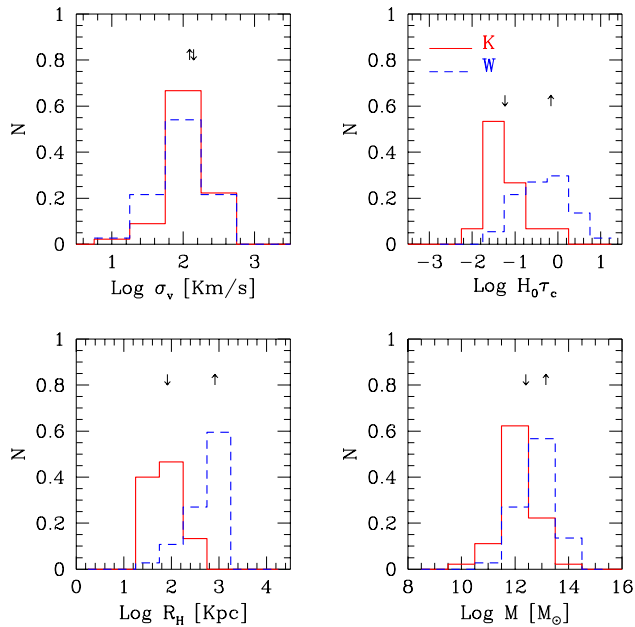}
\caption{ Histograms of dynamical quantities for compact (K, {\it solid line})  and wide (W, {\it broken line}) triplets. Downward arrows indicate  median values for K-triplets, while upward ones for W-triplets. Histograms are normalized to the total number of triplets considered to be physical: 45 for K-triplets and 37 for W-triplets. Notice the overlap in the distributions of K and W-triplets.}
\label{fig:triplets}
\end{figure}

	In Fig.~\ref{fig:triplets}, histograms for the different dynamical quantities are provided,  normalized  to the total number of triplets assumed physical in each of  the above references. 
	We note in passing the existence of overlapping regions between  the K-triplets' and W-triplets' distributions.

\begin{figure}
\epsfxsize=7.5cm
\epsffile{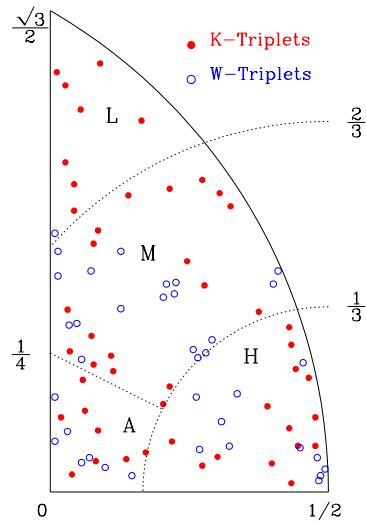}
\vspace{-0.40cm}
\caption{AA-map for triplets.  Karachentsev's triplets ({\it solid dots}) and W-triplets ({\it open circles}) are shown. The areas marked in the map correspond to their different possible configurations: (L) Lagrangian, (H) hierarchical, (A) alignment, and (M) intermediate type. The numbers on the right correspond to the radii of circles centered on (0,1/2).}
\label{fig:agekyan}
\end{figure}

	In addition to the previous kinematical quantities, one may consider also the shape  of the triangle formed by the galaxies in the sky to characterize a triplet; an idea introduced by Agekyan \& Anosova (1968) in terms of an homologous map  (AA-map).
	In this map one normalizes to unity the largest side of the triangle formed by the galaxies in the sky, and the remaining galaxy determines, when a new reference system's axis is set along the longest side, a new set of coordinates which are to be plotted in such diagram.  By symmetry arguments all triangles can be represented inside a particular region; see Fig. \ref{fig:agekyan}.  
	In this AA-map  hierarchical (H) configurations lie at the lower extreme right,  Lagrangian (L) triangles at the upper corner, and alignments (A) or chain-like configurations lie at the bottom of the  diagram.                           

	 Karachentsev's triplets are scattered  more or less randomly in the AA-map, while W-triplets show an absence of L configurations.  
	An important result found by Chernin et al.~\shortcite{chernin-aa} was that projection effects are not responsible for the lack of excess of hierarchical  triplets. This  indicates that K-triplets are absent of many close pairs.  These authors also indicate that W-triplets show only a weak excess of number density in the H-area of the AA-map.

\section{Models for triplet evolution}

\subsection{Galaxy model}

In this work the focus is on the overall dynamics of triplets. Hence we consider it convenient to use a spherically symmetric $N$-body system to model a galaxy. In particular, we have adopted a Plummer sphere. 
	We consider that this galaxy model can represent an  elliptical or, to a first approximation, the  dark halo of a spiral, and can be used to obtain some information about the merging activity of triplets. We have not made any attempt to make an explicit difference between a dark and luminous component.

	The mass enclosed within a radius $r$ and the phase-space density distribution function $f({\cal E})$ for the Plummer model are given, respectively, by:
\begin{equation}\label{eq:plummer}
M(r) = \frac{M \, (r/R_0)^3}{[1+ (r/R_0)^2]^{3/2}}\,, \quad
f({\cal E}) = \frac{24 \sqrt{2} R_0^2}{7 \pi^3\, G^5 M^4}\, |{\cal{E}}|^{7/2}\,,
\end{equation}
where $M$ is the total mass, $R_0$ its scale-length, and $\cal{E}$ the energy per unit mass (Aarseth, H\'enon \& Wielen 1974).
	We constructed each numerical galaxy with $N=3000$ particles, with positions and velocities randomly sampled by an acceptance-rejection technique using the relations in (\ref{eq:plummer}). 
	The $N$-body units used here are such that $G=M=R_0=1$, thus the corresponding unit of time that follows is $t =(R_0^3/GM)^{1/2}$; the total mass of a triplet is $M_{\rm t}=3$. 
	We chose the dynamical time for the galaxy model to be $t_{\rm d}= 2\sqrt{2}(R_0^3/GM)^{1/2}$; i.e., $\sqrt{8}$ in our units. 
	The half-mass radius is $R_{\rm h}\approx 1.3$. 
	Given that the mass distribution of a Plummer sphere falls very rapidly, $\rho \propto r^{-5}$, and about $99$ percent of the enclosed mass is contained within $r \approx 10\,R_0$ we take this value as the extent of our galaxy model.

	In order to compare the results of the $N$-body simulations with those of astronomical systems  we proceeded as follows. We denote by a subscript `$n$' $N$-body values and with an `$a$' the ones in astronomical units.
	We chose the astronomical unit of mass  to be $\mu$ times the approximate mass of the disc of the Galaxy and  $l$ times its disc scale length, respectively:
\begin{equation}\label{eq:units1}
m_{\rm a} =  [ \, 5.1 \times 10^{10} \; \mu \; {\rm M}_\odot \,] \; m_{\rm n}\;, \quad
r_{\rm a} = [ \, 4.5 \; l \; {\rm kpc} \,] \; r_{\rm n} \,.
\end{equation}
These values imply the following time and velocity transformation formulae, respectively: 
\begin{eqnarray}\label{eq:units2}
t_{\rm a} & \, =\, & [\; 20 \; l^{3/2}\,\mu^{-1/2} \;{\rm Myr}\; ] \; t_{\rm n} \,, \nonumber \\
v_{\rm a} & \,= \,& [ \; 221 \; l^{-1/2}\, \mu^{1/2} \;  {\rm km}\, {\rm s}^{-1} \; ]\; v_{\rm n} \,. 
\end{eqnarray} 
If we assume each Plummer sphere represents the Galaxy, with a halo extent of $\approx 135$ kpc and a mass of $\approx 5.5 \times 10^{11}$ M$_\odot$, we have $\mu = 10.78$ and $l=3.01$ as the scaling factors. 
	In particular, these values correspond to Model B of Kuijken \& Dubinski \shortcite{kd} for the Milky Way that has a  mass ratio of dark to luminous matter $\approx 10:1$. This model will constitute our fiducial system to scale our $N$-body results to astronomical ones, unless otherwise stated. Therefore, for our fiducial system, we have:
$r_{\rm a} \approx 13.5 \, r_{\rm n} \, {\rm kpc}$,  
$m_{\rm a} \approx  5.5 \times 10^{11}   \, m_{\rm n}\, {\rm M}_\odot$,  
$t_{\rm a} \approx 32 \, t_{\rm n} \,{\rm Myr}$, and  
$v_{\rm a} \approx 419  \, v_{\rm n} \,{\rm km}\, {\rm s}^{-1}$.

	We followed in isolation the numerical galaxy using a {\sc treecode} (Barnes \& Hernquist 1986) to test its stability. After experimenting with different values, and adopting a set of computational compromises, we decided on  the following numerical parameters for our simulations, both for galaxies in isolation and during the triplet simulations:  tolerance parameter  $\theta=0.9$,   softening parameter $\varepsilon = 0.07\approx R_{\rm h}/20$, time step $\Delta t = 0.1$, and quadrupole corrections to the potential were included. With these parameters energy was conserved to within 0.3 percent during $t/t_{\rm d} \approx 100 $  and to $\approx 0.6$ percent during $t/t_{\rm d} \approx 200$.

\subsection{Initial conditions}

	Theoretically speaking, it will be highly desirable to obtain initial conditions (IC's) from a cosmological scenario for  the formation of triplets; however, this is out of the scope of the present work. 
	None the less,  we have chosen two typical IC's for our triplets, namely:  starting from (1) near `maximum expansion' in a classical cosmology with $\Lambda = 0$, and (2) from virial equilibrium.
	We assume that galaxies are already formed at the moment of starting the simulations. We do not consider the possibility of secondary infall into the region where the triplets will evolve. This is a justifiable assumption in a low density universe similar the one where we appear to live (e.g., Bahcall 1999; Hradecky  et al. 2000),  where major infall of matter has ended at an earlier epoch ($z \sim 1/ \Omega_0$, with $\Omega_0 \sim 0.2$).

	For both types of IC's,  we do not consider the existence of a primordial common envelope of dark matter. This was motivated, on one hand, by the fact that some observations of galaxy clusters and groups tend to suggest that most of their dark matter is associated with the dark haloes of galaxies (e.g., Bahcall, Lubin \& Droman 1995; Puche \& Carignan 1991).
	On the other hand, there are  indications that in binary galaxies their $M/L$-ratios might be  rather small suggesting that large quantities of dark matter in a common halo are unnecessary \cite{honma}. 
	We make the hypothesis that these conditions prevailed once galaxies were already formed in a triplet, a condition that will have to be relaxed in future work where   the effect of a self-gravitating common halo will have to be considered.

	Results from simulations of groups of galaxies indicate that the merging rate is smaller when a common extended massive envelope of dark matter is included (Athanassoula, Makino \& Bosma 1997). In the light of this result,  we would expect that the number of mergers found in our numerical experiments to represent a sort of upper limit. The effect of a primordial concentrated dark halo     is to increase the merging rate, so our results would tend to be a lower limit.

\subsubsection{Turnaround}

 		For IC's simulating the `turnaround' moment for triplets, the centre of mass of each galaxy was  sampled randomly from a uniform spherical density distribution of radius $R_{\rm max}$, which is assumed to be the size of the maximum radius of the density perturbation that led to the formation of a triplet. A small isotropic bulk velocity, derived from a Gaussian with mean zero and one-dimensional velocity dispersion  $\sigma$, was imposed to each galaxy at $t=0$. In a sense, this isotropic $\sigma$ mimics a state of pre-virialization in our simulations (Davis \& Peebles 1977; {\L}okas et al. 1996).

	The numerical values of $R_{\rm max}$ and $\sigma$ were obtained as follows. In the spherical collapse model for the detachment of a density perturbation from the Hubble flow (Gunn \& Gott 1972), the maximum expansion radius at a particular epoch $t$, for a standard cosmology with zero cosmological constant, $\Lambda = 0$, is:
\begin{equation}\label{eq:rmax}
R_{\rm max}(t) = \left(\, \frac{8 G M \, t^2}{\pi^2} \, \right )^{1/3} \;,
\end{equation}
where $M$ represents here the mass at a given epoch.

	We make the hypothesis here that triplets are on the verge of complete collapse for the first time at the present epoch, $t_0$, which, in turn, leads  to a particular  value of $R_{\rm max}$.
	The present age of the universe $t_0$ depends, in the previous cosmology, on the current value of the Hubble constant $H_0$ and the dimensionless density parameter $\Omega_0$. In particular, for an Einstein-de Sitter universe and a low-density universe we have, respectively (Gott et al.  1974): 
\begin{displaymath}
H_0 \,t_0 = \left\{  \begin{array}{lr}
		2/3  &\quad \Omega_0 = 1 \\
		\rightarrow \; 1 + \frac{1}{2}\Omega_0 \ln \Omega_0  & \quad \Omega_0 	\rightarrow 0
		     \end{array}  
	     \right.  \,.
\end{displaymath}
Therefore for $\Omega_0 \approx 0.2$ we have $H_0 t_0 \approx 0.84$ and, using our choice of the Hubble constant,  $t_0 \approx 0.84\, (13\,{\rm Gyr}) \approx 10$ Gyr. 
	In Table~\ref{tab:rmax}, the turnaround radius (in Mpc) of a triplet of Galaxy-like objects ($3\times 5.5\times 10^{11}$ M$_\odot$), for different values of $H_0$ and $\Omega_0$, are given; equation (8) was evaluated at a time $t_0/2$ to obtain $R_{\rm max}$  according to our stated hypothesis.
	Of the plausible values in Table~\ref{tab:rmax}, we have taken $R_{\rm max} = 500$ kpc as our fiducial value for the simulations under this kind of initial conditions.

\begin{table}
\caption[]{Turn-around radius in Mpc}
\label{tab:rmax}
\begin{flushleft}
\begin{tabular}{cccc}
\hline\noalign {\smallskip}
 $H_0$    & $\Omega_0=0$ &  $\Omega_0=0.2$  & $\Omega_0=1$    \\
\noalign{\smallskip}
\hline
\noalign{\smallskip}
50  &  0.84  &  0.74 & 0.64  \\
75  &  0.64  &  0.57 & 0.49  \\
100 &  0.55  &  0.47 & 0.40  \\
\noalign{\smallskip}
\hline
\end{tabular}    
\end{flushleft}
\end{table}

 	Our collapsing triplets are assumed to start with cold initial conditions  satisfying a virial ratio of $2T/|W|=1/4$, as is common in simulations of groups of galaxies (e.g., Barnes 1985). A random velocity, sampled out of a Gaussian distribution with zero mean and 1-D velocity dispersion
\begin{equation}\label{eq-rmax2}
\sigma = \frac{V_0 }{2 \sqrt{3}}\,,  \quad{\rm with} \quad
V_0 = \left ( \frac{3 G M_{\rm t}}{5 R_{\rm max}} \right)^{1/2} \,,
\end{equation}
was assigned as bulk motion to each galaxy.
	Using our fiducial values for $R_{\rm max}$ and the triplet's mass we obtain a value of $\,\sigma \approx 30$ km~s$^{-1}$. 
	This is consistent with the value found by Lake \& Carlberg (1988) for the velocity dispersion of dark matter at the time of turnaround, $\sim 35$ km s$^{-1}$.

\subsubsection{Virial}
	
	Adopting a realistic virial radius from which to sample the position of galaxies is not straightforward, because we are not considering the process of formation of triplets in a cosmological setting. 
	Nevertheless, early point-like simulations made by Peebles (1970) 
suggest that by $\approx 3 t_{\rm c}/2$, with $t_{\rm c}$ being the collapse time scale, a collapsing system has reached an approximate equilibrium state, after having rebounded, with a virial radius  $R_{\rm v} \approx R_{\rm max}/2$ (Gott \& Rees 1975).
	According to Coles \& Lucchin (1995), more recent $N$-body simulations of structure formation in an expanding universe tend to favour this time-scale.

	Therefore, if we make the hypothesis that triplets at the present epoch are virialised, which is suggested by the crossing times of K-triplets, we have that $R_{\rm v} = R_{\rm max}(t_0/3)/2$;  for our fiducial triplet we take $R_{\rm v} = 191$ kpc.
	 The bulk velocities of galaxies were chosen initially so as to satisfy the virial ratio $2T/|W|=1$ exactly.

\subsection{Some computational aspects}

	To obtain the global dynamical properties of triplets, e.g.,  velocity dispersion and mean harmonic radius, we used the centre of each galaxy. We set the most bounded particle of the galaxy model to have 1 percent of its total mass,  set its coordinates to be identically zero in the random realization, and identified this massive particle as the centre of the numerical galaxy.
	This numerical artifact allowed us to use this massive particle to trace the galaxy's path and to compute its bulk velocity by time differentiation.   A similar approach to identify the centre of a numerical galaxy was used by Aguilar \& White (1985).

	We performed 30 triplet simulations for each kind of initial conditions  using the parameters and method discussed above. Each simulation differred   in the initial seed used in the random realization to obtain the positions and velocities of galaxies at $t=0$.
	The collapse time for our fiducial triplet, in $N$-body units,  is $t_{\rm c}= \pi R_{\rm max}^{3/2}(2G M_{\rm t})^{-1/2}\approx 290$ (Gott 1975), and its half-mass free-fall time $t_{\rm ff} = 2^{-1}\pi R_{\rm h}^{3/2}(G M_{\rm t})^{-1/2} \approx 72$ (Mamon 1990); here we have taken $R_{\rm h} = R_{\rm max}/2$.
	The dynamical time for initially virialised triplets is $t_{\rm v}= (2 R_{\rm v})^{3/2}(GM_{\rm t})^{-1/2} \approx 130$. 
	Each triplet was evolved for $t_{\rm n} =321.4$ time units; i.e., $\sim 10$ Gyr.   Energy was conserved to better than $0.5$ percent throughout this time for all simulations. Each run took $\approx 4$ hrs of {\sc cpu} time in a Pentium-II 400MHz PC.

\section{Results and Discussion}

\subsection{Global behaviour}

\begin{figure}
\epsfxsize=8.6cm
\epsffile{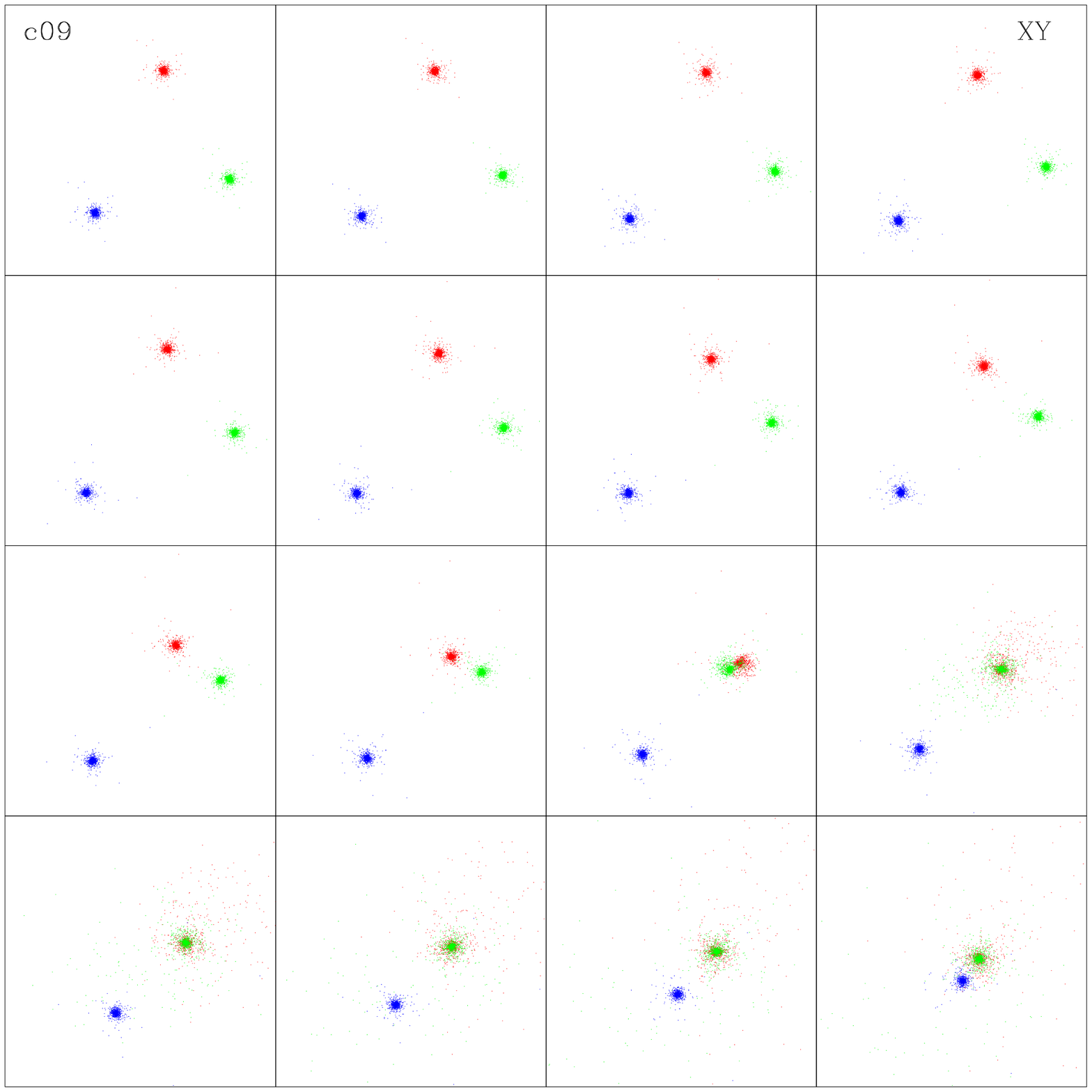}
\caption{Projection on the XY-plane of a particular collapsing triplet simulation. Snapshots are at $\approx 21.4$ time unit ($\approx 0.69$ Gyr) intervals.   Time increases from left to right and from top to bottom. The boxes are $80\times 80$ units ($\approx 1$ Mpc$^2$). Here galaxies do not merge during the whole extent of the simulation ($\sim 10$ Gyr). The present epoch is approximately in the third frame of the first column  from top to bottom.}
\label{fig:c09evol}
\end{figure}

\begin{figure}
\epsfxsize=8.6cm
\epsffile{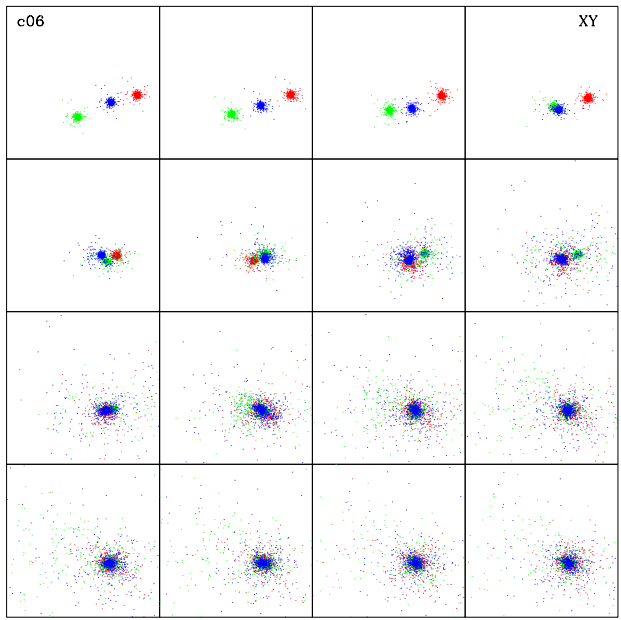}
\caption{Similar to Fig.~\ref{fig:c09evol}, but for virialised triplets.  Here,  the initial conditions lead to a rapid binary formation and a subsequent triple merger. The initial chain-like structure  of the triplet in the sky lasts for $\sim 1$ Gyr. Note the large extent of the halo developed during the merging process. }
\label{fig:c06evol}
\end{figure}

In Figures~\ref{fig:c09evol} and \ref{fig:c06evol} we display the XY-projection of the evolution of two particular triplet simulations starting from maximum expansion.   
These  show some of the general qualitative characteristics found in the other simulations, including those starting in virial equilibrium. For example, a `binary'  galaxy  is in general formed first while the third galaxy `orbits'  around it for some period of time.
	This binary instability is analogous to that found in the 3-body problem (e.g., Valtonen \& Mikkola 1991) and, in this respect,  supports a generalization of this behaviour to triplets of self-gravitating galaxies.

	However, two important qualitative differences between the 3-galaxy and 3-body problem are found: (1) {\it isolated} triplets will eventually merge in their centre of mass provided enough time is given, and (2) no `sling-shot' events were found for these initially bound systems.
	The physical reason for both of these characteristics stems from the fact that  galaxies are able to absorb orbital kinetic energy, and angular momentum,  and transfer it to their internal degrees of freedom. 
	A clear manifestation of the previous is the large extent that matter in a merger can reach; e.g., in Fig.~\ref{fig:c06evol}, the final system can occupy a region of $\sim 1$ Mpc in diameter.
	 If we interpret our galaxy models as representing a spiral, this large region would be occupied primarily by dark matter and form a common halo by the present epoch. This has actually been observed in a  simulation of a triplet of spirals (Aceves 2000b).

 We emphasized above the issue of triplets being isolated because there are some indications that tidal interactions with large-scale structures might disrupt them over a Hubble time (Aceves 2000a). We believe that this may apply more properly to present-day W-triplets due to their large extent.

\subsection{Merging histories}

	We have adopted a  criterion of `loss-of-identity'  (Athanassoula et al. 1997) to determine when a pair of galaxies has merged. 
	Although different merging criteria lead to differences in the number of mergers found at a particular time (e.g., Garc\'{\i}a-G\'omez \& Athanassoula 1993), we consider the adopted criterion well-suited for our purposes,  for it appears more closely related to an observational practice.

	 In particular, we consider that two galaxies to have lost their identity when
\begin{equation}\label{eq:merging}
V_{ij}  = |\bmath{v}_j - \bmath{v}_i | \, <  \, V_{\rm rms}/2 \,, \quad
R_{ij} = |\bmath{r}_j - \bmath{r}_i |\, < \,R_0 /2 \,;
\end{equation}
where $V_{ij}$ and $R_{ij}$ are the relative three-dimensional velocity and distance between galaxies $i$ and $j$, and $V_{\rm rms}$ and $R_0$ are the root-mean-squared velocity and scale radius of a Plummer sphere, respectively. 
	In the numerical implementation of criterion (\ref{eq:merging}) a triple merger required a few time-steps to be detected, since Eq.~\ref{eq:merging} is a pair-wise recipe, but no unreasonable results were obtained.

\begin{figure}
\epsfxsize=8.8cm
\epsffile{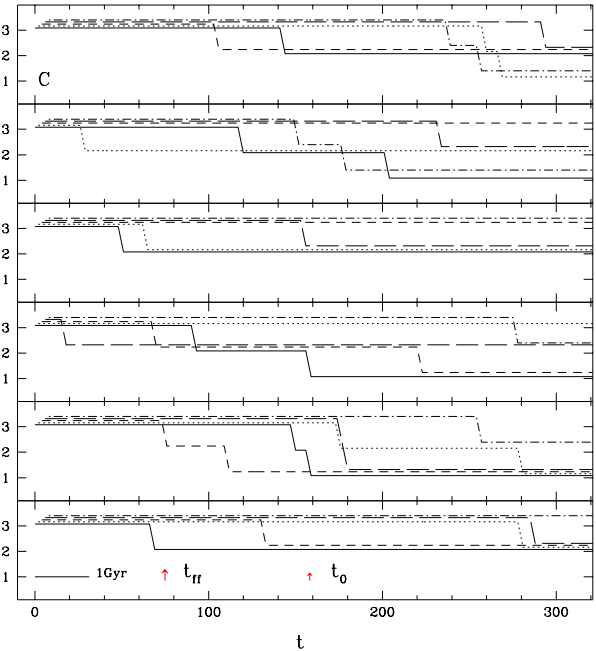}
\caption{Merging histories of thirty collapsing triplets. 
 Time is in $N$-body units and measured from the start of the simulations; indicated are the half-mass free-fall time, $t_{\rm ff}$, and the present epoch, $t_0$, according to our fiducial model. At $t_0$ about 10 per cent of triplets have merged. Merging histories have been slightly displaced horizontally and vertically for better appreciation.}
\label{fig:merghistory-c}
\end{figure}

\begin{figure}
\epsfxsize=8.8cm
\epsffile{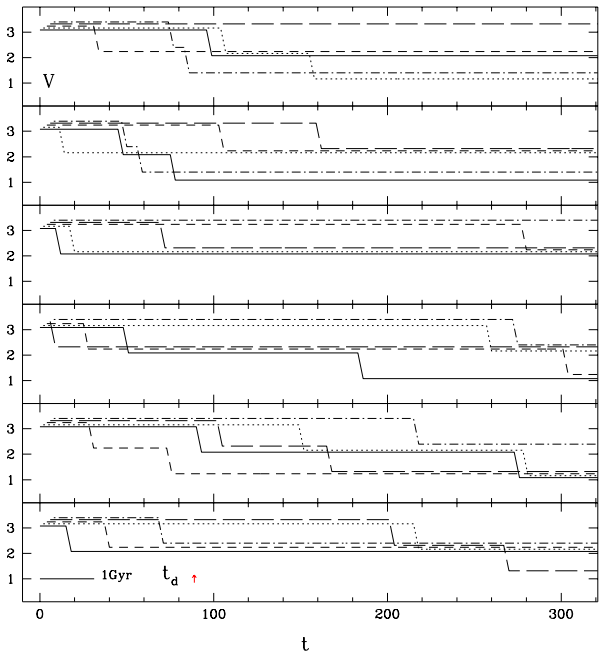}
\caption{Similar to Fig~\ref{fig:merghistory-c}, but for virialised triplets.
The time scale indicated is the dynamical time of the triplet. About 20 percent of triple mergers are found after $\sim 10$ Gyr of evolution.}
\label{fig:merghistory-v}
\end{figure}

In Figs.~\ref{fig:merghistory-c} and \ref{fig:merghistory-v} we show the number of `galaxies' remaining in the triplet at time $t$, both starting from maximum expansion and in virial equilibrium, respectively. A merged pair was counted as an individual `galaxy'.
	We note that the cosmic time,  i.e., the time elapsed since the `big bang', is different for each of the IC's considered; this is because they were assumed to start a different epochs in the evolution of the universe. 
	For IC's starting at maximum expansion the cosmic time is $t_{\rm cos}= t_0/2 + t$, where $t$ is the time span of the simulation (e.g., Ishizawa 1986); analogously, for initially virialised triplets we would have $t_{\rm cos} = t_0 + t$.

	It is interesting to note that by the present epoch, $t_n\approx 160$, only about 10 percent of triplets starting from maximum expansion have merged and about $60$ per cent have not suffered any mergers.
	 At the end of the simulation, i.e.  about 5 Gyr after $t_0$, about 30 percent have managed to survive without any merging.

	 Although there are not many observational studies regarding triplets'  characteristics,  the previous results appear qualitatively consistent with the available observations. For instance, a rapid visual inspection of K-triplets in NASA's database NED does not show strong triple mergers but their luminous components look fairly separated in most of the images, though this may be a consequence of the selection criteria.    One example of a possible triple merger in process is in  Hickson's group 95 (Igl\'esias-P\'aramo \& V\'{\i}lchez 1997), but since this is a group one cannot even strictly compare such `triplet'  with our results, and another is that found by Heidt et al. (1999).

	For initially virialised triplets, 4 ($\approx 15$\%) merge completly in within a dynamical time, while 2 ($\approx 7$\%) do not present any merger and  19 ($\approx 60$\%) only present binary mergers in  $\sim 10$ Gyr. 
	This behaviour is  probably related to the fact that virialised triplets tend to maintain their initial conditions longer, with higher velocity dispersion, hence being less disposed to satisfy the merging criterion (10).
	Obviously, these results would be altered if a primordial common dark halo had been introduced.

	We do not find an overmerging problem for triplets under the kinds of initial conditions considered here, contrary to what is usually found in numerical simulations of compact groups of galaxies (Hickson 1997).  The results obtained also show that there are stable states in the 3-galaxy problem, contrary to what is sometimes believed (Valtonen \& Flynn 2000). 
	We must recognize, however, that since we are not modeling the actual process of formation of a triplet, the above results need to be taken as a first approximation to the problem of merging activity in  triplets.

\subsection{Geometrical properties}
 
	Several authors have used the AA-map to characterize the geometrical properties of triplets and three-body systems (e.g., Chernin et al. 1994;  Hein\"am\"aki et al. 1998). Although this approach  does not provides much information on the dynamical state of triplets, since a triplet initially in the L-area  can migrate toward other regions at later times (e.g., Ivanov, Filistov \& Chernin 1995), can none the less provide us with another means to test our models.

\begin{figure}
\epsfxsize=8.8cm
\epsffile{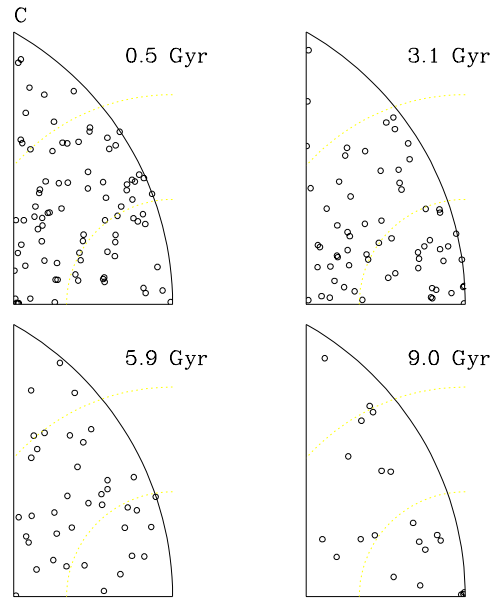}
\vspace{-1cm}
\caption{Evolution of the AA-map for collapsing triplets. Numbers indicate the time elapsed from their initial condition. No significant excess of hierarchical structures are observed at the present epoch ($t\approx 6$ Gyr). Physical hierarchical triplets disappear in this map in a scale of $\sim 1$ Gyr.}
\label{fig:aac-evol}
\end{figure}

\begin{figure}
\epsfxsize=8.8cm
\epsffile{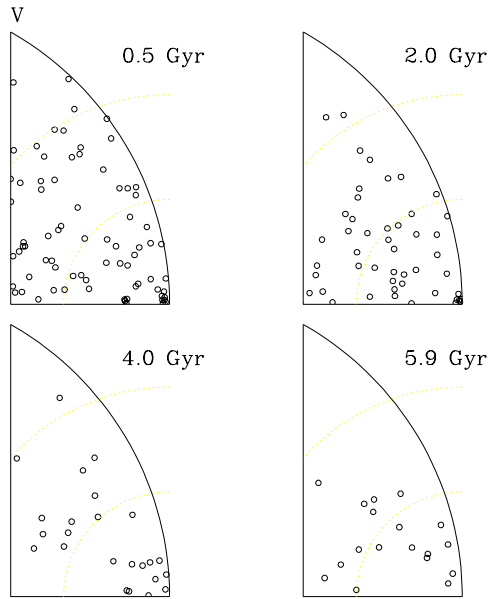}
\vspace{-1cm}
\caption{Similar as in Fig.~\ref{fig:aac-evol}, but for virialised triplets. A  slight excess of hierarchical triplets are observed initially, but later they are no longer present. Merging activity depletes the extreme H-area.}
\label{fig:aav-evol}
\end{figure}

	In Figs.~\ref{fig:aac-evol} and~\ref{fig:aav-evol} we show the AA-map at different times for triplets starting from maximum expansion and virial equilibrium, respectively. Although it is somewhat difficult to assess quantitatively these results and to compare them with the observations of K and W-triplets, in part due to the small number of points for a statistical analysis, it is interesting  to note that the AA-map indicates that no strong hierarchical structures are present at the current epoch ($t\approx 6$ Gyr) for collapsing triplets. On other hand, for initially virialised triplets we obtain a slight excess of triplets in the H-area at $t=0$ that is still present after a few giga-years of evolution.

	 We would like to point out also that it is possible that the lack of many observational triplets in the H-area of the  AA-map may be due to an observational bias. Real galaxies are able to merge and  disappear from the AA-map  at some time.
	Once the haloes of galaxies start to overlap the merging process is rather rapid ($\sim 1$ Gyr), and this will reduce the probability of finding such extreme H-structures in a sky survey.
	If some  W-triplets are truly physical hierarchical triplets one would expect them to disappear from the H-area of the AA-map in the order of a few Gyr.

	The previous results lead us to consider that our models are not inconsistent with the AA-map of observed triplets, both compact and wide. 
	However these results do not allow us to establish a preference for the current dynamical state of triplets. 
	It appears, on other hand, that a large primordial common halo of dark matter is not  required to avoid the the existence of hierarchical triplets,  as suggested by other authors (e.g., Chernin et al. 1994).

\subsection{Dynamical quantities}

In order to compare  our models with observations we have calculated, for both kinds of initial conditions, the one dimensional velocity dispersion, the mean harmonic radius, the dimensionless crossing time, and the virial and median mass along three orthogonal `lines-of-sight' at different times; i.e., we mimic the observation of 90 triplets. 

	In Table \ref{tab:res-col}, we present the median values of the above quantities in $N$-body units for triplets starting at maximum expansion, and in Table \ref{tab:res-virial} the results for initially virialised triplets. The crossing time is expressed in terms of the $N$-body equivalent of our adopted  Hubble time. 	
	The times in Tables~\ref{tab:res-col} and~\ref{tab:res-virial}  correspond to $t_{\rm a}\approx \{0.5,1,...,9\}$ Gyr in our fiducial system of units. 
	In calculating the median of a given quantity at a certain time we have disregarded triplets that have a merger.

\begin{figure}
\epsfxsize=8.7cm
\epsffile{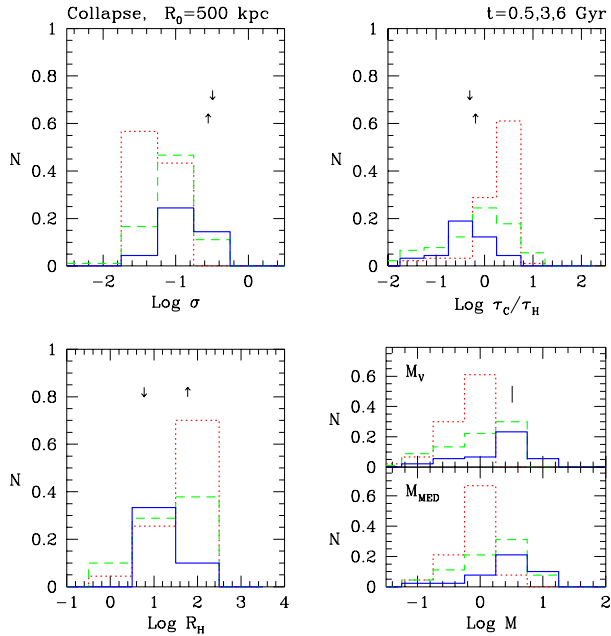}
\caption{Histograms of dynamical quantities for collapsing triplets at different times from their initial state. Line styles correspond to: ({\it dotted}) $t_{\rm a}=0.5$ Gyr; ({\it broken}) $t_{\rm a}= 3$ Gyr; and ({\it solid}) $t_{\rm a}=6$ Gyr. The present epoch is near $t_{\rm a}\approx 6$ Gyr. }
\label{fig:hclpse}
\end{figure}

\begin{figure}
\epsfxsize=8.7cm
\epsffile{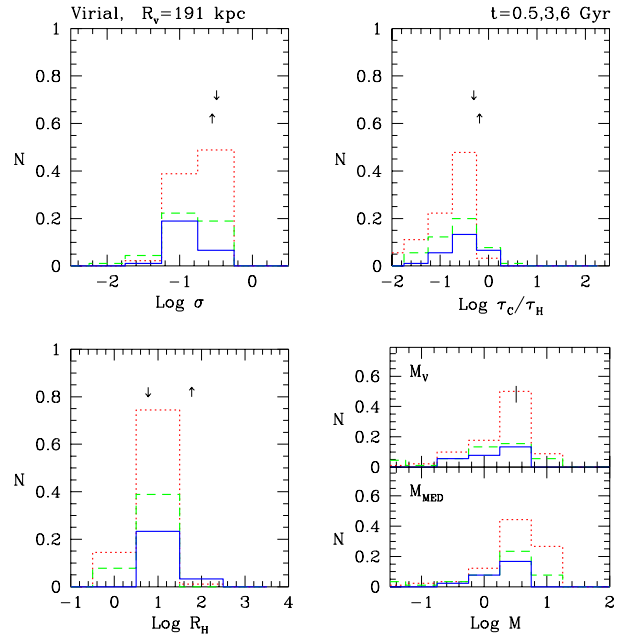}
\caption{Similar as in Fig.~\ref{fig:hclpse}, but for triplets  starting from virial equilibrium.}
\label{fig:hvirial}
\end{figure}

	In Figs.~\ref{fig:hclpse} and~\ref{fig:hvirial}, histograms for the dynamical quantities  at times $t_{\rm a}\approx 0.5,3,6$ Gyr  normalized to the initial number of `observed' triplets are shown   for collapsing and virialised triplets, respectively. This normalization was chosen to show the evolutionary trends in the ensemble of simulations. Downward and upward arrows indicate median observed values for K and W-triplets, respectively. The bar in the mass histograms denotes the total mass of the system in $N$-body units.

\begin{table}
\caption{Turnaround Results}
\label{tab:res-col}
\begin{tabular}{r crr rr  }
\hline
t     & $\sigma$ & $R_{\rm H}$ & $H_0\tau_{\rm c}$ & $M_{\rm V}$ &
$M_{\rm med}$ \\
\hline
17.2  & 0.052 & 39.13 & 2.08 & 0.66 & 0.81 \\   
34.5  & 0.067 & 38.73 & 1.71 & 0.94 & 1.30 \\   
64.6  & 0.080 & 33.93 & 1.37 & 1.16 & 1.71 \\   
94.8  & 0.098 & 31.56 & 0.83 & 1.11 & 1.86 \\   
125.0 & 0.090 & 22.20 & 0.65 & 1.04 & 1.51 \\   
159.4 & 0.125 & 21.40 & 0.43 & 1.96 & 2.07 \\   
185.3 & 0.139 & 19.09 & 0.46 & 2.35 & 3.26 \\   
219.7 & 0.116 & 12.65 & 0.33 & 1.28 & 1.58 \\   
249.9 & 0.141 &  9.37 & 0.16 & 1.16 & 1.71 \\   
282.2 & 0.152 & 14.07 & 0.24 & 1.37 & 2.21 \\   
\hline
\end{tabular}
\end{table}

\begin{table}
\caption{Virialised Results}
\label{tab:res-virial}
\begin{tabular}{r crr rr  }
\hline
t     & $\sigma$ & $R_{\rm H}$ & $H_0\tau_{\rm c}$ & $M_{\rm V}$ &
$M_{\rm med}$ \\
\hline
17.2  & 0.189 & 13.53 & 0.23 & 2.65 & 4.09 \\  
34.5  & 0.152 & 12.73 & 0.18 & 1.62 & 2.23 \\  
64.6  & 0.164 & 14.50 & 0.23 & 2.05 & 2.83 \\  
94.8  & 0.159 & 15.30 & 0.23 & 1.14 & 2.51 \\  
125.0 & 0.135 & 15.96 & 0.29 & 2.22 & 2.78 \\  
159.4 & 0.120 & 14.42 & 0.34 & 1.40 & 2.81 \\  
185.3 & 0.126 & 13.40 & 0.33 & 1.80 & 3.18 \\  
219.7 & 0.176 & 12.71 & 0.17 & 1.87 & 2.29 \\  
249.9 & 0.161 &  8.72 & 0.13 & 1.05 & 2.65 \\  
282.2 & 0.187 & 13.89 & 0.24 & 3.53 & 3.84 \\  
\hline
\end{tabular}
\end{table}

In general, as expected,  the median properties of the virialised triplets tend to evolve much slower than those for the collapsing ones, which is perhaps most noticeable in  the small change of the median  velocity dispersion (Table~\ref{tab:res-virial}). 
	On the other hand, a significant evolution toward the establishment of a virial equilibrium among the translational bulk degrees of freedom of galaxies is observed for collapsing triplets. This is noticeable as their virial mass estimate approaches their true mass. 
	However, because galaxies are self-gravitating systems, part of their  mutual orbital kinetic energy is transformed into internal kinetic energy and, thus, no complete virial equilibrium in their translational bulk degrees of freedom can be achieved. 
	A virial equilibrium, but in the complex mixture of the  internal and bulk degrees of freedom of galaxies, tends to be established during their interactions.

	From Tables~\ref{tab:res-col} and \ref{tab:res-virial}, we observe that no overestimates of the median virial mass  are obtained at any time, except at the end of the simulations for virialised triplets. 
	Overestimates were found in few of our numerical experiments, but only for brief intervals of time when using the median mass estimator. At most a $\approx 20$ per cent overestimate was found for virialised IC's  and $\approx 10$ per cent for collapsing triplets. 
	After examining each run, we found that several triplets were a factor  at most of about $4$ higher than the true mass of the triplet, but this happened only along a particular line-of-sight while along the other two an underestimation always occurred. 
	These overestimates happened  before the galaxies overlapped of the galaxies; i.e., when galaxies still behaved somewhat  as point-like particles and not as self-gravitating systems.
	These results are  in contrast to others that indicate that overestimates of even a few orders of magnitudes may be obtained in triplets.

	The results of Table~\ref{tab:res-col}  indicate that  triplets arriving at the present epoch ($t_{\rm n}\approx 160$) from maximum expansion  have  their median virial masses underestimated by $\approx 35$ percent, with a somewhat better agreement when using the median mass estimator. The physical reason behind this underestimate, in these truly bounded systems, is the already-mentioned  transfer of orbital to internal energy. 
	This means, in observational practice, that the use of the bulk velocities  of galaxies --even if we had three-dimensional information--  has the intrinsic propensity to underestimate  the total mass of the system. A similar conclusion was reached by Barnes (1985) in his study of small groups of galaxies. 
	For initially virialised triplets (Table~\ref{tab:res-virial}) the median mass estimator tends to provide a better estimate, though it may overestimate in some cases, by less than $\approx 30$ percent.

	Overall, we consider the median mass estimator to perform better than the virial mass estimator in these low-multiplicity systems. The accuracy of the mass estimate depends in an important way on the dynamical status of the system.  To derive a time-varying expression for the median or virial mass estimator is not the purpose here. See however  Dolgachev \& Chernin (1997) for a study developed to take this into account using point-like particles as  a galaxy model.

	At the present epoch, $t_{\rm n}\approx 160$, our models predict that triplets starting from maximum expansion should have a median $R_{\rm H} \approx 290$ kpc  and $H_0\tau_{\rm c} \approx 0.43\,$. 
	When comparing these values with the properties of  K-triplets (Table 1),  we find them higher by about a factor of 4 and 10, respectively.
	Therefore, it appears that our models yield results that are difficult to reconcile with the observations.
	This situation is worse when we note that a median of    $\sigma \approx 52$ km s$^{-1}$ is predicted, a value which is about half of the observed one for K-triplets. A maximum of $\sigma \approx 64$ km s$^{-1}$ is obtained at the end of the simulations. 
	For virialised IC's, the maximum median is $\sigma \approx 80$ km s$^{-1}$, occurring at $t=0$. Afterwards, the median $\sigma$ tends to decrease somewhat due to the transfer of orbital to internal energy during the interactions of galaxies.

	The previous results, taken at face value, tend to rule out the models presented for triplets with either kind of initial conditions considered.
	However, there are some observational matters and numerical aspects  that indicate that our models might not be an unreasonable scenario for the global evolution of triplets. In $\S$\ref{sec:evolution} we will establish which kind of initial conditions we favour the most.

	On other hand, when comparing our results to observations we are assuming implicitly that K-triplets form an homogeneous sample, both in luminosity ($\sim$ mass) and morphological type, but as can be seen from the properties of K-triplets~\cite{kara89} this is not the case.
	In this respect our ensemble of numerical triplets forms, by construction, a homogeneous `catalog'.
 	A mass-spectrum, for example, can have  effects on the  numerical values obtained for $\sigma$ and $R_{\rm H}$ that are not easy to estimate. 
	This motivated us to estimate the velocity dispersion of K-triplets weighted by the luminosity of their member galaxies in order to compare it with our numerical results. Weighting by mass appears to be a more uncertain procedure. 	The median weighted velocity dispersion obtained was $\sigma\approx 90$ km~s$^{-1}$ (about 25 per cent  lower the value in Table \ref{tab:triplets}), which in turn is closer to our results.

  	Different selection criteria can yield significantly different results for the dynamical quantities in galaxy systems. For example, in a recently-announced catalog of small galaxy groups by  Makarov \& Karachentsev (2000), using   physically motivated selection criterion, the following median values, with quartiles, for triplets are reported: (a) velocity dispersion $41_{-18}^{+19}$ km~s$^{-1}$, (b)  mean harmonic radius $191_{-88}^{+157}$ kpc, and dimensionless crossing time $0.15_{-0.09}^{+0.16}\,$ (using $H_0=70$ km~s$^{-1}$~Mpc$^{-1}$). These authors provide a harmonic radius value not corrected for projection effects (i.e. lower by a factor of $\pi/2$ than when using Eq. \ref{eq:rh}) and use the crossing time definition of Gott, Wrixon \& Wannier (1973) (Makarov, {\it private communication}). The latter definition is $t_{\rm c}=3 R_{\rm H} (5^{3/2}\sigma)^{-1}$, that, when comparing with Eq.~\ref{eq:tcross}, yields $\tau_{\rm c} \approx 4 t_{\rm c}\,$. 
	Therefore, comparing  our results at $\approx t_0$ for collapsing triplets (Table~3)  with the values found by Makarov \& Karachentsev  a  much better agreement is found.
	Furthermore, we note that while all the previous comparisons were done among the median values, we find that $\approx 10$ per cent of collapsing triplets have $\sigma \sim 100$ km~s$^{-1}$ at $\approx t_0$ and small RH's ($\S$\ref{sec:evolution}).

	Part of the disagreement of our median results with, for example, the   observed velocity dispersion of K-triplets may be somewhat alleviated if we use a larger mass as our fiducial galaxy model, but it does  not satisfactorily solve the problem.  
	Zaritsky et al. (1997) after examining the distribution and kinematics of satellites around  spiral galaxies conclude that the halo masses of spirals are in the excess of $2 \times 10^{12}$ M$_\odot$ within $\approx 200$ kpc. 
	Using this type of galaxies, the scaling factors $\{l,\mu\}$ in Eq. \ref{eq:units1} are $\{4.4,39.2 \}$. 
	From this we obtain that $v_{\rm a}= 660 v_{\rm n}$ km~s$^{-1}$, which leads to a median $\sigma \approx 80$ km~s$^{-1}$ at $\approx t_0$ for collapsing triplets. This new scaling provides a much better agreement with the velocity dispersion observed in K-triplets. 
	However, the discrepancy with observations moves now to the $R_{\rm H}$ median value that results in  $\approx 425$ kpc, i.e.,  about an order of magnitude larger than the observed value and marginally consistent with the results of Makarov \& Karachentsev (1999). The time scale does not change by much. 
	One may try to force a better agreement by selecting a galaxy with a mass of $\approx 10^{12}$ M$_\odot$ within $\approx 100$ kpc, but this appears in conflict with the fact that not many galaxies satisfying these conditions, e.g., large ellipticals, exist in K-triplets. 
	To properly address  these scaling matters, more realistic simulations including a mass-spectrum would be required.

	The models considered here do not properly apply to present day observed W-triplets. A better model for these would be, perhaps, to assume that they are in the early phases of their turnaround. With this assumption   $R_{\rm max} \sim 1$ Mpc and $\sigma \sim 30$ km~s$^{-1}$. The latter scenario has been considered briefly elsewhere  (Aceves 1999a), where it was found that we will have to wait for about another Hubble time for W-triplets to reach a compact configuration similar to the one  of K-triplets. See also Dolgachev \& Chernin (1997). However,  the current large  $\sigma \sim 100$ km~s$^{-1}$ for W-triplets does not seems to favour such possibility.

	We note that W-triplets appear to be a less homogeneous sample than K-triplets, for they include triplets with a very low harmonic radius similar to those of K-triplets (Fig.~1).
	It is not the purpose here to make an analysis of the selection criteria used to compile W-triplets, but we point out that some of them would appear more properly classified as K-triplets.
	The overlapping in the RH-distribution is interesting in itself, however, since  some triplets classified as W-triplets may constitute a link between those that are in the early phases of collapse  and those that are in the  more advanced phase.

	Larger samples of triplets and more detail studies of their properties are  required to build a more solid observational database for future comparisons with theoretical models. Efforts toward expanding the sample of triplets, for example, in the southern sky are presently being done by several groups (e.g., Karachentseva \& Karachentsev 2000).

\subsection{Evolutionary Trends}\label{sec:evolution}

An important question to investigate is how K-triplets have  attained their present configuration and dynamical properties. From the previous results,  it appears that triplets starting from a diffuse configuration, followed by a collapse, is a probable scenario.

	Perhaps one of the best methods to discern the evolution of a dynamical system is to look into its phase-space structure. 
	However this is not possible in galaxy systems since we do not have yet enough information to construct a snapshot of their physical phase-space. 
	None the less, we may construct a pseudo phase-plane for triplets  by using their velocity dispersion and mean harmonic radius, compare them with the numerical results, and see if some evolutionary trends are discernible.

\begin{figure}
\epsfxsize=8.5cm
\epsffile{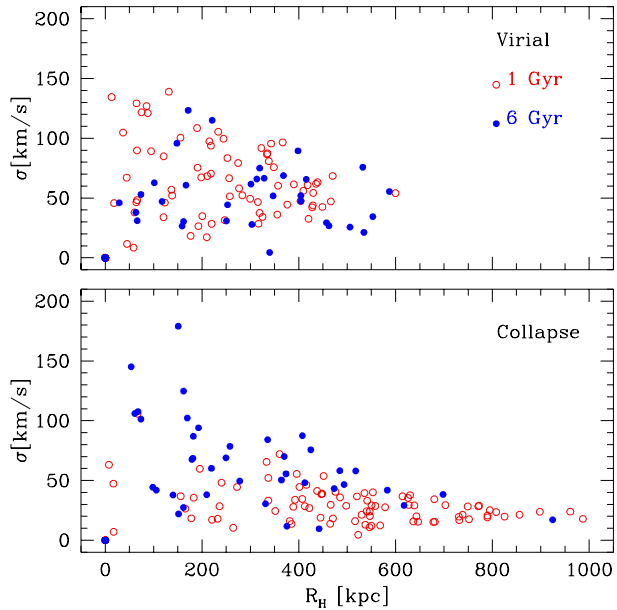}
\caption{Pseudo phase-plane for triplets under both types of initial conditions considered and at different times during the simulations. The present epoch is at $\approx 6$ Gyr for collapsing triplets. }
\label{fig:phspace}
\end{figure}

\begin{figure}
\epsfxsize=8.5cm
\epsffile{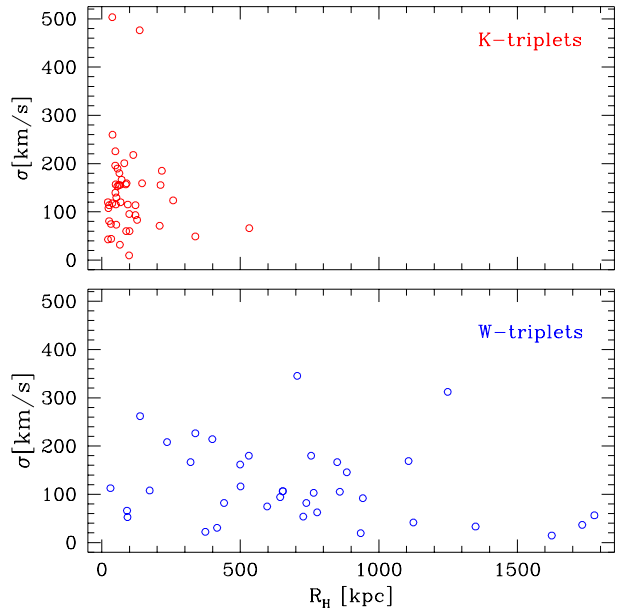}
\caption{Pseudo phase-plane for observational triplets considered to be physical systems, both compact and wide. Notice the change of  scale from that in   Fig.~\ref{fig:phspace}, and  how some W-triplets occupy part of the region of K-triplets.}
\label{fig:phspacetri}
\end{figure}

In Fig.~\ref{fig:phspace},  we show the pseudo phase-plane (hereafter, phase-plane for brevity) for our numerical triplets at different times spanned from $t=0$, both for collapsing and virialised initial conditions.
	In Fig.~\ref{fig:phspacetri} the phase-planes of K and W-triplets are shown.
	Again it is difficult to assert some conclusions from a comparison of the phase-planes, but it is interesting to note that some tendencies exist. 
	For example, K-triplets show, on one hand, the general trend followed by  systems that had suffered a collapse out of a much wider configuration: a  decrease in their $\sigma$ as $R_{\rm H}$ increases.
	This behaviour is similar to that exhibited by the numerical triplets that have started from maximum expansion in our simulations.
	On the other hand, W-triplets have a phase-plane that resembles qualitatively systems that are on the verge of suffering a gravitational collapse.

	From the phase-plane of collapsing triplets shown in Fig.~\ref{fig:phspace}, we observe that about 10 percent of them have dynamical quantities ($\sigma$ and $R_{\rm H}$) similar to the median observed in K-triplets; c.f. Table~1. This is also observed in some triplets with virial IC's at $t_0$.

	In general, we consider that the results and data shown in these phase-planes, along with earlier discussions,  suggest that: K-triples are probably the most extreme cases of triplets that started from a diffuse configuration, attained their current dynamical properties via a gravitational collapse process, have  manage to survive without suffering major  mergers, and are near virial equilibrium in their bulk degrees of freedom.

\section{Summary and Conclusions}

	A simplified $N$-body model for the dynamical evolution of triplets, using self-consistent galaxies, has been proposed. Two types of initial conditions, assuming a classical cosmology with $\Lambda = 0$, have been investigated, namely  `maximum expansion' and virial equilibrium.
	Triplets starting from maximum expansion were imposed the condition that, assuming all their mass to be distributed homogeneously, would collapse at the present epoch, and those in virial equilibrium  to be in  such state now.
	No primordial common dark halo was included in our simulations.

	The main results of this work are as follows:

\begin{enumerate}

\item   The use of self-consistent galaxies to study the dynamics of triplets is of fundamental importance. No `sling-shot' events occur during a triple  interaction in an initially bound system, in contrast with what is a common event in the 3-body problem.

\item	A rather low number of triple mergers  ($\approx 10$ per cent) are expected to occur at the present epoch for systems starting from maximum expansion. This appears to be in concordance with observations.  Initially virialised systems yield $\approx 5$ per cent of triple mergers in $\sim 10$ Gyr of evolution. Therefore, no overmerging problem is found for triplets. This contrasts to what it is usually found in numerical simulations of small compact groups.

\item Under both types of initial conditions considered, we find that the 3-galaxy problem  has stable states; i.e., there are certain  conditions that do not lead to a triple merger even after $\sim 10$ Gyr of evolution.

\item  The geometrical properties of numerical triplets, as indicated by their  homologous AA-map, are not inconsistent with those of observed triplets, both compact and wide. 
	Physical triplets in the extreme hierarchical region of the AA-map would suffer a rapid dynamical evolution toward a binary merger leading to a depletion of triplets in  such region, since they would no longer would be counted as triplets in an observational survey.

\item	Median values of the dynamical parameters for triplets that have started as a diffuse system and are collapsing for the first time at the present epoch  are not inconsistent with the data of Karachentsev's compact triplets.  A much better agreement is reached with recent data on triplets. We find that about $\approx 10$ percent of the simulated triplets reproduce well the K-triplets median dynamical quantities at the present epoch.

\item   The median of the virial mass estimator does not in general overestimate  mass, but, on the contrary, it underestimates it. For triplets in an advance stage of collapse at the present epoch an underestimate of  $\approx 35$ percent is found. 
	Therefore if K-triplets are truly bound physical systems, with no primordial common dark halo,  their  $M/L$-ratios are being underestimated by the same factor. Otherwise, the underestimate may be larger.
 	The median mass estimator appears as a somewhat better mass estimator for triplets than the virial mass estimator.

\item 	The plane $\sigma-R_{\rm H}$ of numerical triplets suggests that K-triplets have acquired their present configuration through a clustering process, and lead us to conclude that they are the triplets with the most extreme values in the literature, i.e., lowest harmonic radius and highest velocity dispersion.  	Some dynamical properties of K-triplets are also consistent with a virial equilibrium state in their bulk degrees of freedom at the present time.

\item Our results suggest that the presence of a primordial, large, massive, common dark halo might not be a strong requirement to account for the dynamical evolution of triplets, but they cannot exclude it.

\end{enumerate}

In order to better compare with observations, future models should include, e.g., a mass-spectrum for galaxies, consider different morphological types, and investigate the role of a primordial common dark halo  in a cosmological setting.

\section*{Acknowledgments} 
The Spanish Ministry of Foreign Affairs (MUTIS program) and M\'exico's CONACyT   (Proyecto I35546-E) are thanked for financial support at different stages of this work.  We thank greatly Gary Mamon for fruitful discussions on the issue of initial conditions.  Jack Sulentic and H\'ector Vel\'azquez are also thanked for comments on this work, and Michael Richer for help with the english writing.



\bsp
\label{lastpage}

\end{document}